\documentclass[%
 jmp,
 amsmath,
 amssymb,
 preprint,%
author-year,%
author-numerical,
]{revtex4-1}
\usepackage[dvipdfmx]{graphicx}
\usepackage{dcolumn}
\usepackage{bm}
\usepackage{subfigure}
\usepackage{sidecap}

\def\vc(#1){\mbox{\boldmath $#1$}}

\def\abs(#1){\left|#1\right|}

\def\lb{\left(}
\def\rb{\right)}

\def\barr(#1){\begin{array}{#1}}
\def\earr{\end{array}}

\def\roman(#1){\mbox{\rm #1}}

\def\beq{\begin{eqnarray}}
\def\eeq{\end{eqnarray}}

\def\lb{\left(}
\def\rb{\right)}

\def\barr(#1){\begin{array}{#1}}
\def\earr{\end{array}}
\def\roman(#1){\mbox{\rm #1}}
\def\beq(#1){\begin{eqnarray}\label{#1}}
\def\eeq{\end{eqnarray}}
\def\bfig(#1){\begin{figure}[htbp]\label{fig:#1}}
\def\efig{\end{figure}}

\def\vc(#1){\mbox{\boldmath $#1$}}

\begin{document}

\title[
Stochastic modeling on fragmentation process over lifetime and its dynamical scaling law of fragment distribution
]{
Stochastic modeling on fragmentation process over lifetime and its dynamical scaling law of fragment distribution
}

\author{Shin-ichi Ito}
\author{Satoshi Yukawa}
\affiliation{Department of Earth and Space Science, Graduate School of Science, Osaka University, Toyonaka 560-0043, Osaka, Japan}

\date{\today}
\begin{abstract}
We propose a stochastic model of a fragmentation process, developed by taking into account fragment lifetime as a function of their size based on the Gibrat process. If lifetime is determined by a power function of fragment size, numerical results indicate that size distributions at different times can be collapsed into a single time-invariant curve by scaling size by average fragment size (i.e., the distribution obeys the dynamical scaling law). If lifetime is determined by a logarithmic function of fragment size, the distribution does not obey the scaling law. The necessary and sufficient condition that the scaling law is obeyed is obtained by a scaling analysis of the master equation.
\end{abstract}
\maketitle

\section{Introduction}
Fragmentation of material is a commonly observed process and has been investigated experimentally
\cite{0295-5075-25-6-004, JPSJ.61.3474, Austin1976,PhysRevLett.78.1444}
 and theoretically
 \cite{Montroll1940, PhysRevE.54.4293, Ziff1986, Gilvarry1961,
  PhysRevE.86.016113, krapivsky2003shattering, 0305-4470-26-22-011,
  ben2000fragmentation, krapivsky1994scaling}. 
Statistical quantities are useful for understanding the characteristic features of such phenomena. The size distribution of a fragment is a statistical quantity characterizing the fragmentation process. For example, when a material is fractured by an external force such as in impact fragmentation, size distribution is known to show a power-law or log-normal distribution
\cite{PhysRevLett.78.1444, PhysRevE.86.016113,
  JPSJ.61.3474}.
  In these types of fragmentation process, the subdivision process of fragments is stopped immediately. In this case, the characteristic quantities of power law or log-normal law are determined by the properties of the material. In particular, the log-normal distribution can be simply explained by the Gibrat process
\cite{crow1988lognormal},
 a discrete-time stochastic process.
 We assume that a fragment always breaks into two random pieces.
At the time $t=n$, there is a fragment of size
$S_{n}$. In next time step $t=n+1$, the fragment breaks,
and its size becomes
$S_{n+1}=r_{n}\times S_{n}=S_{0}\Pi_{i=0}^{n}r_{i}$, where $r_{n}$ is
the random dividing ratio of the fragment
and $S_0$ is the initial size of the fragment.
The logarithm of $S_{n}$, $\log S_{n}$, is distributed normally at larger $n$ as per the central limit
theorem, and the average $E_{n}$ and variance $V^{2}_{n}$
of the fragment size can be derived as $E_{n}=nm$ and
$V^{2}_{n}=ns^{2}$, where $m$ and $s^{2}$ denote the average and 
variance of the distribution of $\log r_{n}$. 
As a result, $S_{n}$ is
distributed log-normally, and the functional form of the probability
density function $P\lb S, n\rb$ is described by:
\begin{equation}
P\lb S, n\rb dS=\dfrac{dS}{\sqrt{2\pi V^{2}_{n}}S} \mbox{\rm exp}\left[
  -\dfrac{\lb\mbox{\rm log}\lb S\slash S_{0}\rb -E_{n}\rb^{2}}{2
    V^{2}_{n}}\right] \enspace.
\end{equation}

The fragmentation process may also be caused by an internal force such as a desiccation stress\cite{0295-5075-25-6-004}.
There are few studies that concentrate on the size distribution in this scenario. Subdivision develops more slowly than in the case of collisional fragmentation. In this situation, the characteristic values of the size distribution are determined not only by the material properties but also the external environment.
S.~Ito and S~.Yukawa \cite{1209.6114}
simulated a continuum model of the fracture process of drying viscoelastic thin paste and showed that the time series of the area distributions of a fragment can be collapsed into a single master curve by scaling with the average value. This scaling law is called a ``dynamical scaling law'' and cannot be expressed in terms of the original Gibrat process because the resultant lognormal distribution includes two parameters of the average  $E_{n}$ and the variance $V_{n}$. Generally there is no relationship between $E_n$ and $V_n$; the lognormal distribution therefore can not be scaled by the average alone.

In this article, we investigate the dynamical scaling law in the fragmentation process of desiccation using a stochastic model based on the Gibrat process. While the original Gibrat process clearly cannot describe the dynamical scaling law, understanding the statistical properties of a fragmentation process is straightforward. In this article, we therefore model the fragmentation process of drying paste by a stochastic process that extends the Gibrat process. While in the original Gibrat process the effect of desiccation is not taken into account, we here incorporate it as the ``lifetime'' of the fragments. Thus, the breaking event does not occur simultaneously but is instead dependent on the lifetime of each fragment.

The organization of this paper is as follows: In Sec.~\ref{sec:model}, the proposed stochastic model based on the Gibrat process is introduced. This model takes into account the lifetime of the fragment, which depends on fragment size. In Sec.~\ref{sec:result}, numerical and theoretical results are shown. The time evolutions of size distributions and their average values are investigated and the dynamical scaling property is demonstrated and explained theoretically. Section~\ref{sec:summary} presents conclusions and discussion. In App.~\ref{sec:appa} we calculate the lifetime of the fragments based on elastic theory.

\section{\label{sec:model}Fragmentation process incorporating fragment lifetime}

In the original Gibrat process, each fragment breaks after a constant time interval. When considering fragmenting material, the time interval is generally not constant, but rather expected to depend on fragment size (expressed as length, area, or volume). The time interval between two successive breaking events of a fragment is termed its ``lifetime'' and is a function of fragment size. Introducing lifetime into the original Gibrat process yields a stochastic process referred to as ``modified Gibrat process'' in the following.

\begin{figure}[htbp]
  \includegraphics[width=1.0\linewidth]{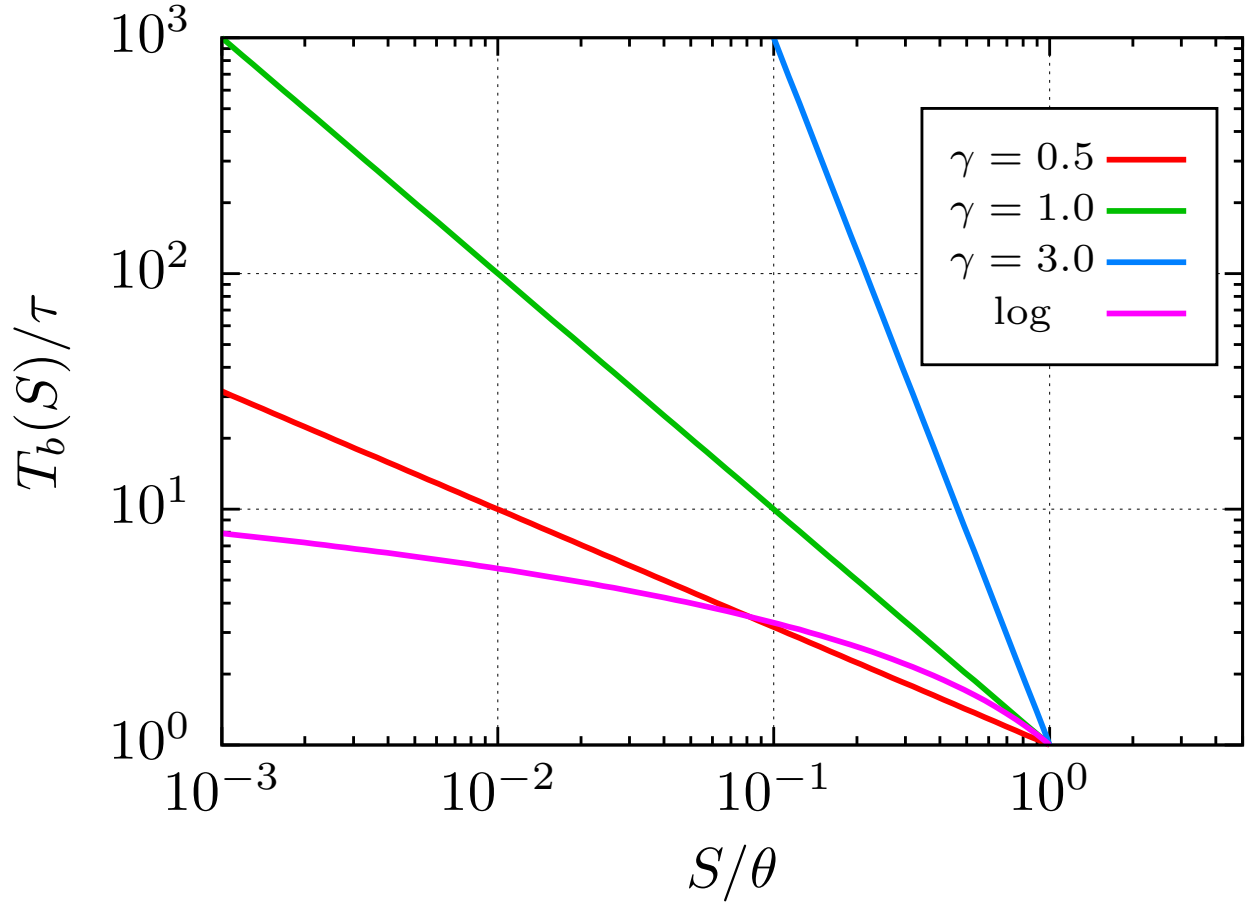}
  \caption{
    (color online) The functional form of lifetime $T_{b}$.
    Red, green and blue lines correspond to cases of $\gamma=0.5$,
    $1.0$ and $3.0$ of $T_{b}\lb S\rb=\tau \lb S\slash \theta\rb^{-\gamma}$, respectively.  
The magenta line corresponds to the case of $T_{b}\lb S\rb=\tau \lb 1+\mbox{\rm log}\theta-\mbox{\rm log}S\rb$.
  }
  \label{fig:tb}
\end{figure}

In this paper, we assume that lifetime is a function of fragment
size, $S$, which we denote as $T_{b}\lb S\rb$.
We use a power function and a logarithmic function as the functional
form of the lifetime: 
\begin{equation}
T_{b}\lb S\rb=\tau \lb
\dfrac{S}{\theta}\rb^{-\gamma}
\label{eq:lifetimep}
\end{equation}
and
\begin{equation}
T_{b}\lb S\rb=\tau \lb 1+\mbox{\rm
  log}\theta-\mbox{\rm log}S\rb \enspace,
\label{eq:lifetimel}
\end{equation}
where $\tau$ and $\theta$ are the
characteristic time and size, respectively(see Fig.~\ref{fig:tb}). 
In these expressions, we choose the functional forms to give $\tau$ for 
the initial size $\theta$. 
The relationship between lifetime and fragment size in the example fragmentation process is discussed in App.~\ref{sec:appa}.

\begin{figure}[htbp]
  \includegraphics[width=1.0\linewidth]{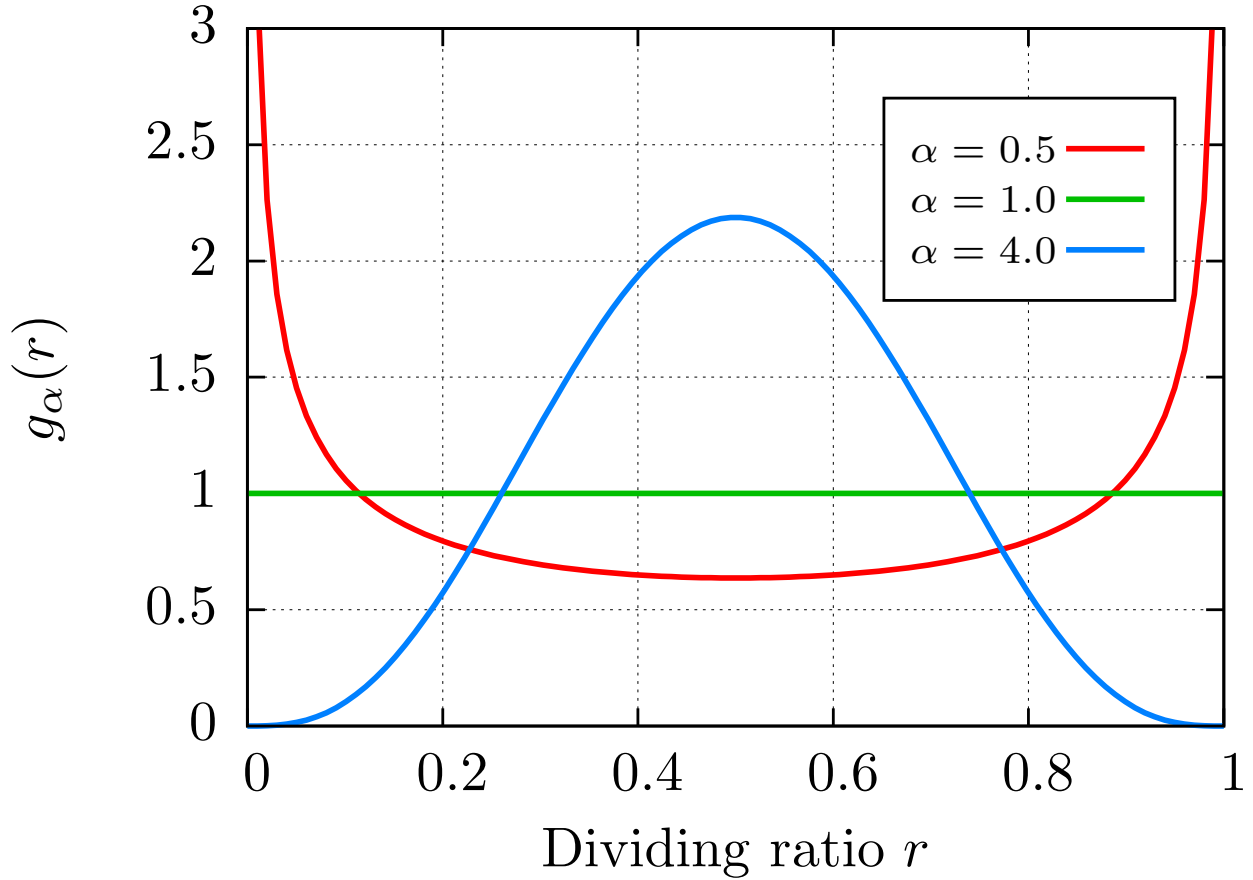}
  \caption{
    (color online) Beta distributions used
    as dividing ratio distribution.
    Red, green and blue lines correspond to $\alpha=0.5$,
    $1.0$ and $4.0$, respectively.  
  }
  \label{fig:prob}
\end{figure}

In the modified Gibrat process, we must introduce a probability
density function of fragment dividing ratio.
In this paper,
the beta distribution,
$g_{\alpha}(r)=r^{\alpha-1}\left(1-r\right)^{\alpha-1}/B(\alpha,\alpha)$,
is used as the probability density function (see Fig.~\ref{fig:prob}), 
where $r$ is a dividing ratio and $B(\alpha,\alpha)$ is a beta function used for 
normalization of the distribution.
For simplicity, we restrict the domain of $g$ to the open interval $(0,1)$ in the following analysis.
We require
symmetry relative to the dividing ratio $1\slash 2$ and controllability of
the variance of the distribution by a single parameter $\alpha$, that is, a
uniform distribution for $\alpha =1$ and a Gaussian-like distribution
for much larger $\alpha$. 

The modified Gibrat process consists of the following procedure (Fig. 3). We start with a single specimen with size $S_0=\theta$ and lifetime $T_{b}\left(S_{0}\right)$ determined by $S_{0}$. After the lifetime has passed, the specimen breaks into two fragments with sizes $S_1 = r S_0$ and $S_2 = (1-r)S_0$, where the ratio $r$ is randomly chosen from the distribution $g_\alpha (r)$. The procedure is repeated for each new fragments. 
\begin{figure}[htbp]
  \begin{center}
    \includegraphics[width=1.0\linewidth]{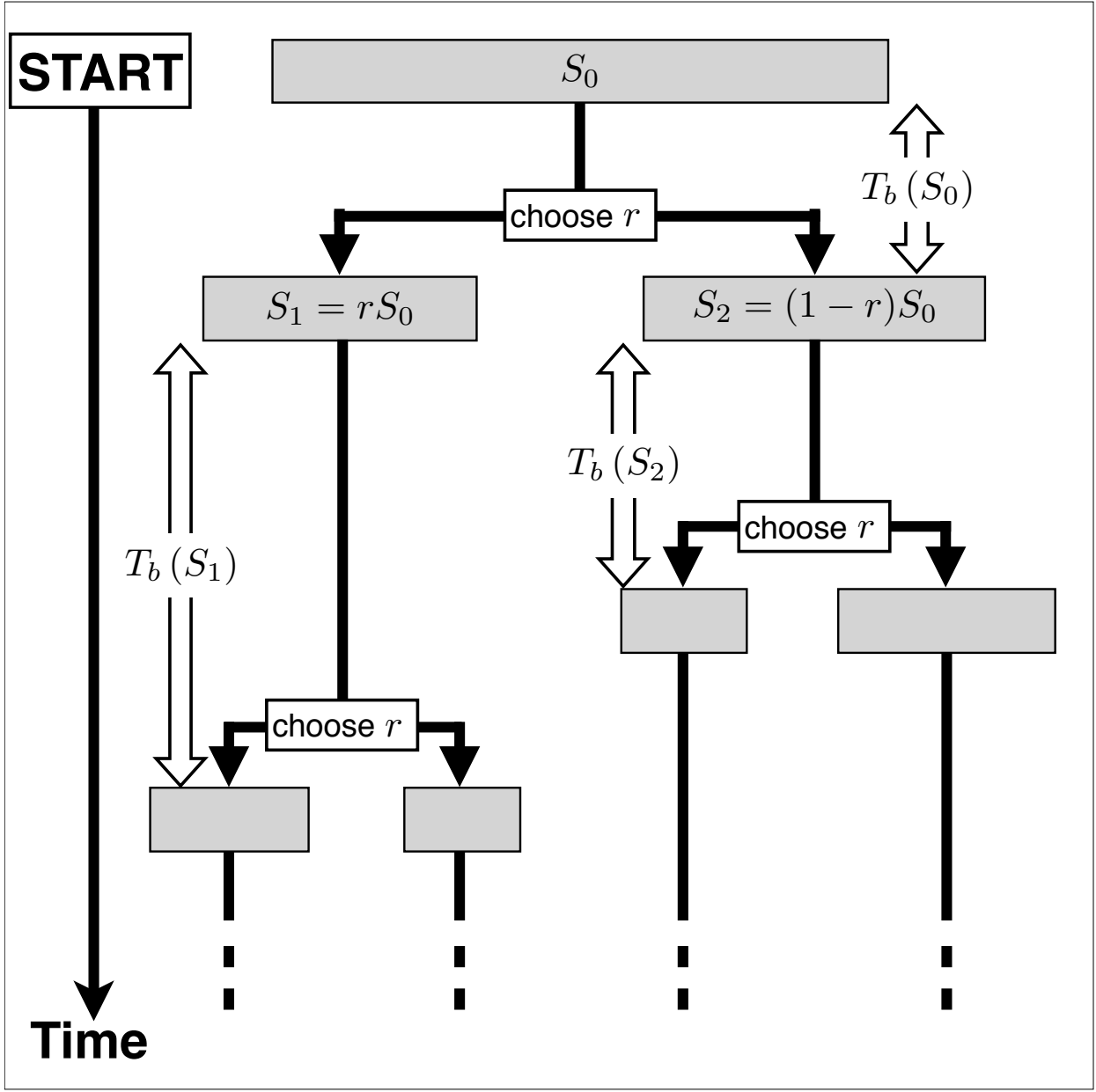}
    \caption{
      Schematic flow diagram of the modified Gibrat process. The initial fragment has size $S_{0}$ and lifetime $T_{b}\left(S_{0}\right)$ determined by $S_{0}$. When $T_{b}\left(S_{0}\right)$ has passed, the fragment breaks into two pieces following dividing ratio ${r}$, which is chosen from a probability density function $g_\alpha\lb r\rb$, resulting in two fragments with sizes $S_{1}=rS_{0}$ and $S_{2}=\left(1-r\right)S_{0}$. The procedure is repeated for the new fragments with size $S_{1}$ and lifetime $T_{b}\left(S_{1}\right)$, and size $S_{2}$ and lifetime $T_{b}\left(S_{2}\right)$.
    }
    \label{fig:tree}
  \end{center}
\end{figure}

\section{\label{sec:result}Results}

\begin{figure}[htbp]
	\begin{center}
	\includegraphics[width=1.0\linewidth]{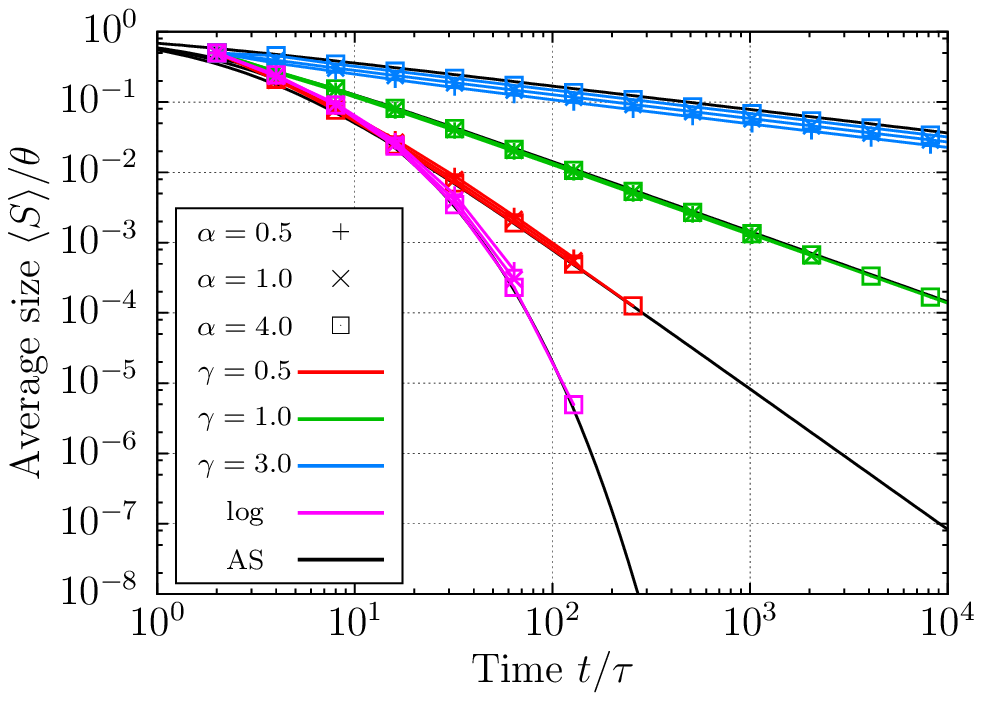}
	\caption{
          (color online)
Time evolutions of average size. Difference in color indicates difference of the functional form of lifetime. Symbols represent the parameter $\alpha$. Black lines denoted as “AS” are approximate solutions from Eq.~\eqref{chap4-2}. 
          }
	\label{fig:averagesize}
	\end{center}
\end{figure}

Figure~\ref{fig:averagesize} shows the time evolution of average size $\langle S \rangle$ of the fragments. This does not depend on $\alpha$ (the functional form of $g_{\alpha}(r)$). In addition, the time series of the average size can be fitted by $t^{-1\slash\gamma}$ if lifetime is a power function with an exponent $-\gamma$ (Eq.~\eqref{eq:lifetimep}). If lifetime is logarithmic (Eq.~\eqref{eq:lifetimel}), the time series of average size can be fitted by $\mbox{\rm exp}\lb
-C\sqrt{t}\rb$, where $C$ is a positive fitting parameter.

This behavior can be understood by the following way.
When the $n$-th breaking event is finished, the size of the
fragment is given by $S_{n}=S_{0}\Pi_{i=0}^{n-1} r_{i}$. 
At this time, the elapsed time is the summation of the lifetime and is
given by $t_{n}=\sum_{i=0}^{n-1}T_{b}\lb S_{i}\rb$.  
Substituting $S_{i}$ with $Z_{i} = \log S_{i}$, that is, $S_{i} = e^{Z_{i}}$, 
and writing $T_{b}(S_{i}) = W(Z_{i})$, we obtain 
\begin{equation}
t_{n}  = \sum_{i=0}^{n-1} T_{b}(S_{i}) = \sum_{i=0}^{n-1} W(Z_{i})
\enspace .
\end{equation}
As $ \left\lvert Z_{i+1} - Z_{i}\right\rvert = \left\lvert \log S_{i+1}/S_{i}\right\rvert = \left\lvert\log r_{i+1} \right\rvert \simeq  \log 2 < 1$, this summation is approximated by the following integral: 
\begin{equation}
t_{n} \simeq \int_{Z_{0}}^{Z_{n}}\dfrac{dZ}{-\log 2} W(Z)=-\dfrac{1}{\log 2}\int_{S_{0}}^{S_{n}} \dfrac{dS}{ S} T_{b}(S)
\label{chap4-1}
\enspace .
\end{equation}
Assuming the discrete variables $(t_{n}, S_{n})$ to be continuous variables $(t,S)$, we obtain
\begin{equation}
t =-\dfrac{1}{\log 2}\int_{S_{0}}^{S} \dfrac{dS}{ S} T_{b}(S)
\enspace .
\end{equation}
Differentiating the above equation with respect to $S$, we obtain the following differential equation: 
\begin{equation}
\dfrac{dS}{dt}=-\log 2 \dfrac{S}{T_{b} (S)} \enspace . 
\label{chap4-2}
\end{equation}  
Originally $S$ is a stochastic variable. Here, however, we should treat it as the average 
$\langle S \rangle$ because of the approximation $\log r_{i+1} \simeq - \log 2$.
Solving the equation with an
appropriate initial condition of size, an asymptotic solution can be
obtained:
\begin{equation}
\langle S \rangle\sim t^{-1\slash\gamma} 
\end{equation}
for  $T_{b}\lb S \rb=\tau \lb S\slash\theta\rb^{-\gamma}$ and 
\begin{equation}
\langle S \rangle\sim
\exp \lb -\text{const.}\sqrt{t} \rb
\end{equation}
for $T_{b}\lb S \rb=\tau\lb 1+\log\theta-\log S\rb$.
These approximated solutions describe the numerical data well and are shown in Fig.~\ref{fig:averagesize}.

\begin{figure*}[htbp]
  \begin{center}
    \includegraphics[width=1.0\linewidth]{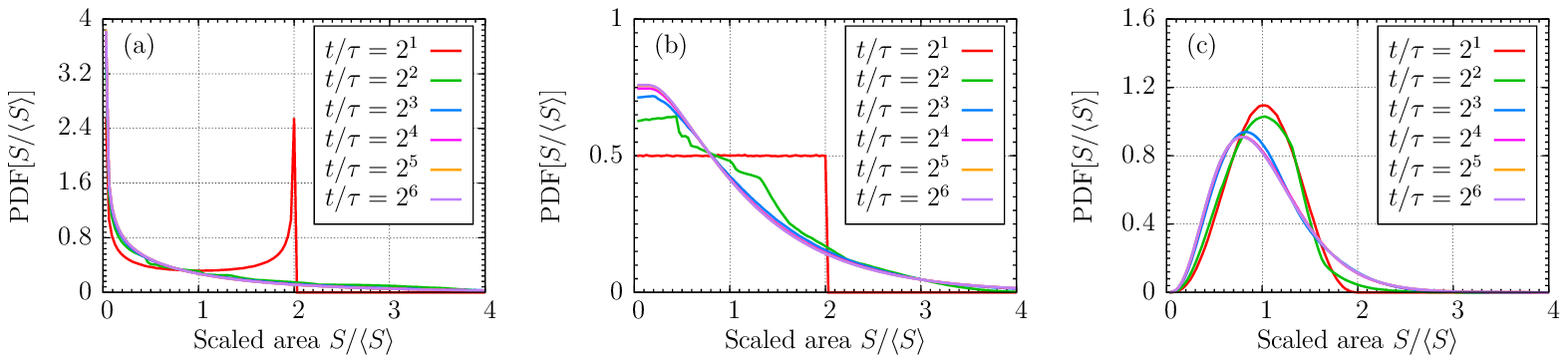}
    \caption{(color online) Time series of probability density
      functions (PDFs) of scaled size $S\slash\langle S \rangle$ with 
      lifetime given by $T_{b}\lb S \rb=\tau \lb
      S\slash\theta\rb^{-\gamma}$ and $\gamma=0.5$.  Different colors correspond to different times.
Figures (a), (b) and (c), indicate the PDFs using $\alpha=0.5$ ,$1.0$ and $4.0$, respectively.
     	}
	\label{fig:pdfg0.5}
	\end{center}
\end{figure*}

\begin{figure*}[htbp]
  \begin{center}
    \includegraphics[width=1.0\linewidth]{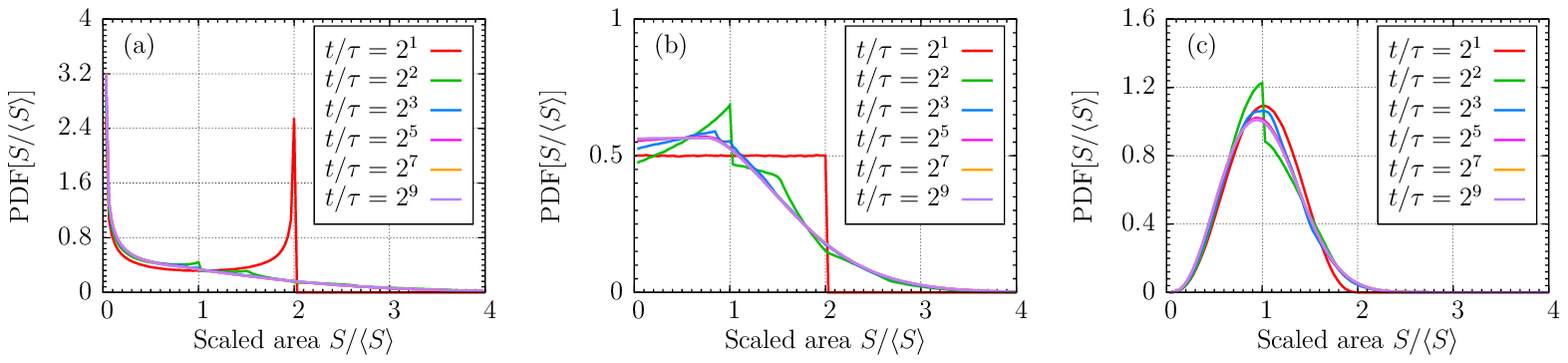}
    \caption{(color online) Time series of probability density
      functions (PDFs) of scaled size $S\slash\langle S \rangle$ with 
      lifetime given by $T_{b}\lb S \rb=\tau \lb
      S\slash\theta\rb^{-\gamma}$ and $\gamma=1.0$.  Different colors correspond to different times.
Figures (a), (b) and (c), indicate the PDFs using $\alpha=0.5$ ,$1.0$ and $4.0$, respectively.
     	}
	\label{fig:pdfg1.0}
	\end{center}
\end{figure*}

\begin{figure*}[htbp]
  \begin{center}
    \includegraphics[width=1.0\linewidth]{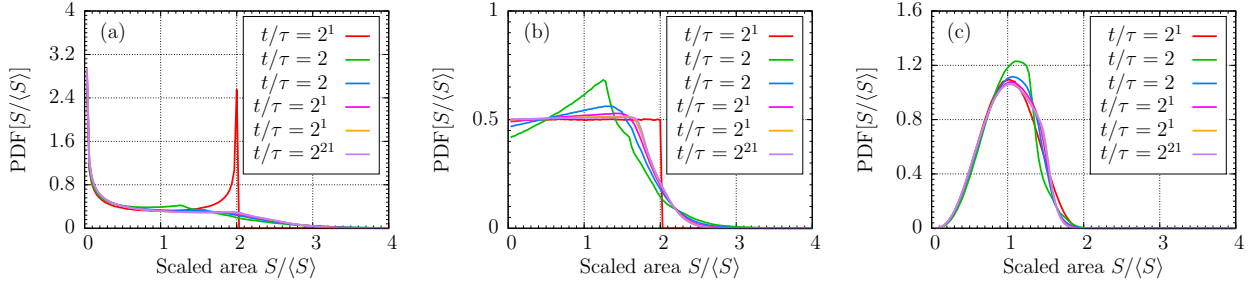}
    \caption{(color online) Time series of probability density
      functions (PDFs) of scaled size $S\slash\langle S \rangle$ with
       lifetime given by $T_{b}\lb S \rb=\tau \lb
      S\slash\theta\rb^{-\gamma}$ and $\gamma=3.0$.  Different colors correspond to different times.
Figures (a), (b) and (c), indicate the PDFs using $\alpha=0.5$ ,$1.0$ and $4.0$, respectively.
     	}
	\label{fig:pdfg3.0}
	\end{center}
\end{figure*}

Figures~\ref{fig:pdfg0.5}, \ref{fig:pdfg1.0} and \ref{fig:pdfg3.0} show the time evolutions of size 
distributions in the case of $T_{b}(S)=\tau(S\slash\theta)^{-\gamma}$ with $\gamma=0.5$, $1.0$ and $3.0$, respectively.
(a), (b) and (c) of each figure correspond to the case of $\alpha=0.5$, $1.0$ and $4.0$, respectively.
Horizontal axes are scaled by average size.
At the first break ($t=2^{1}\tau$), each distribution takes the obvious form
that corresponds to $g_{\alpha}$.
Distributions converge to specific shapes that are independent of time. These results demonstrate the existence of a dynamical scaling law, in that the time-dependent size distribution function
$P(S, t)$ can be described by a single argument-scaling function $\tilde{P}(X)$ with $X = S \slash \langle S \rangle$:
\begin{equation}
\label{chap4-3}
P\lb S,t\rb dS=\tilde{P}\lb X\rb dX
\quad \text{ with } \quad X\equiv \frac{S}{\langle S \rangle}.
\end{equation}

\begin{figure*}[htbp]
	\begin{center}
	\includegraphics[width=1.0\linewidth]{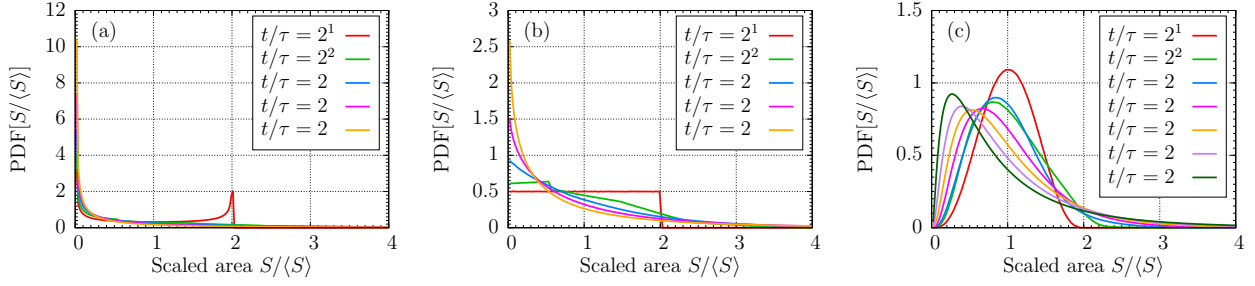}
	\caption{
	(color online)
	Time series of probability density functions (PDFs) of scaled size $S\slash\langle S \rangle$ with lifetime given by $T_{b}\lb S \rb=\tau \lb1+\mbox{\rm log}\theta-\mbox{\rm log}S\rb$.
	Different colors correspond to different times.
	Figures (a), (b) and (c) correspond to cases of $\alpha=0.5$, $1.0$ and $4.0$, respectively.
	}
	\label{fig:pdflog}
	\end{center}
\end{figure*}
Figure~\ref{fig:pdflog} shows the time evolutions of size distributions in the case of $T_{b}\lb S \rb=\tau\lb1+\log\theta-\log S\rb$.
(a), (b) and (c) correspond to cases of $\alpha=0.5$, $1.0$ and $4.0$, respectively.
At the first break ($t=2^{1}\tau$), 
each distribution again takes the obvious form. Size distribution functions grow divergently at smaller scaled size, $S\slash \langle S\rangle < 1$.
As a consequence, the time series of size distributions do not converge into a master curve. Thus in this case, there is no dynamical-scaling property.

The scaling property may be explored theoretically by markovianizing the modified Gibrat process.
The master equation is given by
\begin{multline}
\label{chap5-1}
	\dfrac{\partial P\lb S,t\rb}{\partial t}=-\frac{P\lb S,t\rb}{T_{b}\lb S \rb} \\
	+\int_{0}^{\infty}dS'\int_{0}^{1}dr\,g_{\alpha}\lb r\rb \delta\lb rS'-S\rb\frac{P\lb S',t\rb}{T_{b}\lb S' \rb}\enspace .
\end{multline}
Assuming a scaling transformation $P\lb S, t \rb  \rightarrow  P\lb \beta S,\eta t \rb$,
we require that the master equation is invariant. This is a necessary and sufficient condition and is given by
\begin{equation}
\label{chap5-2}
\frac{\eta}{T_{b}\lb \beta S\rb} =\frac{1}{T_{b}\lb S\rb}  \enspace . 
\end{equation}
This implies 
\begin{equation}
T_{b}\lb S \rb \propto S^{-\gamma} 
\end{equation}
where $\gamma\equiv -T_{b}'\lb1\rb\slash T_{b}\lb1\rb$ and 
\begin{equation}
\eta=\beta^{-\gamma}\enspace .
\end{equation}
It is found that the dynamical scaling law is only valid if lifetime is determined a power function and is independent of
$g_{\alpha}\lb r\rb$.  
For non-zero $\gamma$, the scaling relation 
\begin{equation}
\label{chap5-3}
P( S,t) dS= \beta P(\beta S, \eta t) dS = \eta^{-1\slash \gamma} P ( \eta^{-1\slash \gamma} S,\eta t)  dS.
\end{equation}
provides the scaling law by assuming $\eta=1\slash t$. We obtain the dynamical scaling form as
\begin{equation}
\label{chap5-4}
P(S,t) dS=t^{1\slash \gamma}P (  t^{1\slash \gamma} S,1) dS.
\end{equation}

The average value $\langle S \rangle=\int_{0}^{\infty}dS\,SP\lb S,t\rb$ can also be discussed. Substituting the scaling relation Eq.~\eqref{chap5-3} into the definition of the average,
$\langle S \rangle$ is given by
\begin{equation}
\label{chap5-5}
\langle S \rangle =\mathcal{F}( t )=\beta \int_{0}^{\infty}\!dS\,SP( \beta S,\eta t).
\end{equation}
Replacing $\beta S$ with $x$ at $t=1$, Eq.~\eqref{chap5-5} becomes:
\begin{align*}
\mathcal{F}(1)&=\beta \int_{0}^{\infty}\!dS\,SP( \beta S,\eta ) \\
&=\beta^{-1} \int_{0}^{\infty}\!dx\,xP(  x,\eta )= \beta^{-1}\mathcal{F}( \eta )\\
&\Leftrightarrow \mathcal{F}( \eta) = \mathcal{F}( 1 )\eta^{-1\slash \gamma}.
\end{align*}
Thus we find that $\langle S \rangle=\mathcal{F}\left( 1 \right) t^{-1\slash \gamma}$.
The dynamical scaling law may then be derived from Eq.~\eqref{chap5-4} and $\langle S \rangle =\mathcal{F}\left( 1 \right) t^{-1\slash \gamma}$:
\begin{align*}
	P(S,t) dS&=\mathcal{F}\left( 1 \right)P \left( \frac{S\mathcal{F}\left( 1 \right)}{\langle S \rangle},1\right) \dfrac{dS}{\langle S \rangle}\\
		&=\tilde{P}( X) dX \quad \text{   with   } \quad X\equiv \dfrac{S}{\langle S \rangle}.
\end{align*}

\section{\label{sec:summary}Conclusion and discussion}

We designed a stochastic process incorporating fragment lifetime based on the Gibrat process. If lifetime is determined by a power function of fragment size, the time series of size distributions was found to be collapsible by scaling size using average size. The scaling law is obeyed universally and is independent of dividing ratio distribution. The modified Gibrat process has a strong non-Markov property, and number of stochastic process variables increases with time because of ongoing fragmentation. By representing stochastic variables by average size, an ordinary differential equation of average size can be obtained, and the solution becomes a good approximation of the time evolution of average size in the modified Gibrat process. We obtained a master equation by approximating the modified Gibrat process as a Poisson process. Scaling analysis proved the existence of a dynamical scaling law if lifetime is determined by a power function of fragment size. In the general case, it is however difficult to solve the master equation analytically.

Appendix~\ref{sec:appa} presents the case of a quasi-two-dimensional desiccating fragment and discusses the relationship between fragment area and lifetime $T_{b}$.
Briefly, $T_{b}$ is a function of the square of the characteristic length of the fragment and therefore is described as a function of fragment area. It is dependent on the inverse function of ``desiccation stress'' $f(t)$.
Desiccation stress is the time-increasing negative hydrostatic pressure due to the evaporation of liquid content and corresponds to the specific drying process. Actual lifetime thus reflects the property of the desiccation process.
The power (logarithmic) function $T_{b}$  corresponds to the power (exponential) function $f$.
In previous studies of drying crack patterns\cite{1209.6114,Kitsunezaki1999,Otsuki2005,Nishimoto2007,Kitsunezaki2010,Nag2010}, 
a linear or exponential function is often used as desiccation stress $f$.
Results in the present study yield the dynamical scaling law if $f$ is determined by a linear function, but not in the case of an exponential function. The functional form of $f$ depends on the drying process as desiccation stress has a strong correlation with the amount of liquid content. In experiments using natural drying crack patterns (i.e. no artificial drying procedure is applied), the dynamical scaling law appears to be observed. While it may therefore be expected that $f$ is a power function in the natural drying case, it can not be confirmed the functional form due a lack of experimental studies. We surmise that it is a linear function.

The presented stochastic process was developed based on a fragmentation process of drying crack patterns. However, using lifetimes corresponding to specific phenomena, it is applicable to various fragmentation processes. In addition, it may be possible to classify various fragmentation processes by measuring fragment lifetime.

\section*{Acknowledgement}

The authors are grateful to A. Nakahara, S. Kitsunezaki, T. Ooshida and M. Otsuki for constructive discussions. SI would like to thank K. Kanazawa for useful discussions of stochastic processes. The numerical calculations in this work were carried out on SR16000 at YITP in Kyoto University and the facilities of the Supercomputer Center, Institute for Solid State Physics, University of Tokyo. SI acknowledges the support of a Grant-in-Aid for JSPS Fellows. This work was partly supported by Grant-in-Aid for Scientific Research (C) No. 22540387 from JSPS, Japan.

\appendix

\section{\label{sec:appa} lifetime of a dying thin layer}

In this section, the lifetime of a drying fragment is discussed analytically. We derive the relationship between lifetime and fragment size from elastic theory, considering the fragmentation process of viscoelastic paste by a desiccation process.

We consider a quasi-two-dimensional viscoelastic continuum attached to a flat base. It is assumed that the motion of the material is over-damped due to a strong viscosity and that the drying speed is sufficiently slower than the typical velocity scale of the material dynamics. Because it adheres to the base, the material is also subject to a resistance force proportional to the displacement field. Let $\vc(u)=(u_{x},u_{y})$ and
$\vc(\sigma)=\left( \begin{array}{cc} \sigma_{xx}&\sigma_{xy}\\ \sigma_{yx}&\sigma_{yy} \end{array}\right)$ denote the displacement field and the stress tensor, respectively. In the over-damped case, elastic and resistance forces are balanced. Then the motion equation becomes
\begin{equation}
\label{chap2o1}
\nabla\cdot\vc(\sigma)=k\vc(u),
\end{equation}
where $k$ is a proportional constant of the resistance force, which can be evaluated as $\mu\slash H^{2} $ with the second Lam\'e constant $\mu
$ and the thickness of the paste $H$.\cite{1209.6114}
From the above assumptions, the constitutive equation of the stress can be approximated as a linear elastic equation. In addition, an increasing stress is introduced into the diagonal term of the stress tensor as an effect of desiccation. The constitutive equation is given by
\begin{equation}
\label{chap2o2}
\vc(\sigma)=\lb \lambda\mbox{\rm tr}\lb \nabla \vc(u)\rb+F_{0}f\lb \frac{t}{\tau}\rb \rb \vc(E) + \mu\lb \nabla \vc(u) + \lb \nabla \vc(u)\rb^{T}\rb.
\end{equation}
where $\lambda$ and $\mu$ are the first and second Lam\'e constant, respectively, and 
$\vc(E)$ is an unit tensor.
$F_{0}f\lb t\slash\tau\rb$ is the effect of the desiccation. 
This term works as a negative pressure physically caused by the evaporation of the inner liquid.
$F_{0}$ and $\tau$ denote characteristic stress and time, respectively.
$f\lb z\rb$ is an arbitrary increasing function.  
Due to this negative pressure, tension develops within the material. This tension conflicts with the resistance force due to adhesion. As a consequence, stress concentration appears in the material.

It is assumed that the fragment starts to break when the inner stress exceeds a threshold value $\sigma_{Y}$. 
The ``lifetime'' is defined by the interval between the time when the fragment is created and the time when it starts to break, , and can be calculated using Eqs.~\eqref{chap2o1} and \eqref{chap2o2}.
In the present case we do not need to calculate the general and exact solution, but rather a rough estimation.
When fragment size is characterized by $L^{2}$ with a characteristic length $L$, the characteristic scale of changing the displacement field and stress field can also be described by $L$. Lifetime can then also be calculated by means of dimensional analysis. 
Let $U$ and $\mathfrak{S}$ denote the characteristic scale of the displacement field and stress field, respectively.
Eqs.~\eqref{chap2o1} and \eqref{chap2o2} can be rewritten as follows:
\begin{align}
\label{chap2o3}
	-\frac{\mathfrak{S}}{L}& =kU,\\
\label{chap2o4}
	\mathfrak{S}&=\lb\lambda+2\mu\rb \frac{U}{L} + F_{0}f\lb \frac{t}{\tau}\rb.
\end{align}

Eliminating $U$ from Eqs.~\eqref{chap2o3} and \eqref{chap2o4}, we obtain $\mathfrak{S}$ as 
\begin{equation*}
	\mathfrak{S}=\frac{F_{0}f\lb t\slash \tau\rb}{1+\lb \lambda_{D}\slash L\rb^{2}},
\end{equation*}
where $\lambda_{D}=\sqrt{\lb\lambda+2\mu\rb\slash k} = \sqrt{(\lambda + 2\mu)\slash \mu} H$.
This quantity represents the length scale of the stress, which characterizes how deep the influence of the boundary penetrates into the interior of the material.
Because the lifetime is defined as the time when $\mathfrak{S}$ exceeds $\sigma_{Y}$ from $t=0$, lifetime $T_{b}$ is determined as follows:
\begin{equation}
\sigma_{Y}=\dfrac{F_{0}f\lb T_{b}\slash \tau\rb}{1+\lb \lambda_{D}\slash L\rb^{2}}
	\Leftrightarrow T_{b}=\tau f^{-1}\lb \dfrac{\sigma_{Y}}{F_{0}} \left\{ 1+  \left(\dfrac{\lambda_{D}}{L}\right)^{2}\right\}\rb,
\end{equation}
where $f^{-1}\lb y\rb$ is an inverse function of $f\lb z\rb$.
$T_{b}$ shows asymptotic behavior, as $L$ becomes smaller as the fragmentation is proceeding. For the case of $L \gg \lambda_{D}$ (i.e the earlier state of fragmentation), the second term in the argument of $f^{-1}$ can therefore be ignored.
As $L$ decreases asymptotically, the second term becomes the dominant contribution to $T_{b}$.
The behavior of $T_{b}$ is thus as follows:
\begin{equation*}
	T_{b}\lb L\rb\sim\begin{cases}
		\tau f^{-1}\lb \dfrac{\sigma_{Y}}{F_{0}}\rb =\text{const.} & \qquad \text{if} \;\; L\gg \lambda_{D} \\
		&\\
		\tau f^{-1}\lb \dfrac{\sigma_{Y}}{F_{0}}\cdot\dfrac{\lambda_{D}^{2}}{L^{2}}\rb & \qquad \text{if} \;\; L\ll \lambda_{D}.
\end{cases}
\end{equation*}
This behavior is consistent with the physical interpretation of $\lambda_{D}$: if the system size $L$ is greater than $\lambda_{D}$, the inner stress increases without being effected by the boundary conditions, and $T_{b}$ becomes a constant value depending on $\sigma_{Y}$. In the opposite case, the inner stress is affected by the influence of the boundary conditions - in other words, the desiccation stress may decrease effectively because of the boundary. In this case, $T_{b}$ becomes much larger than in the previous case.

The case of a constant lifetime corresponds to the Gibrat process,
thus the fragmentation process in the early stage follows this process.
After the early stage(when $L > \lambda_{D}$), the lifetime depends on the characteristic length of the fragment $L$. 
In this stage, we can expect dynamical scaling for the power-function lifetime.

The validity of the above dimensional analysis can be confirmed in the simple case of a disk-shaped fragment. In this situation, we can obtain the exact solution of Eqs~\eqref{chap2o1} and \eqref{chap2o2}.
Assuming a disk fragment with a radius $R$, it is assumed for initial conditions that $\vc(u)=0$, $\vc(\sigma)=0$, and that normal stress on the boundary is zero. The lifetime can be calculated exactly and is given by
\begin{align*}
	T_{b}(R)&=\tau f^{-1}\lb \frac{\sigma_{Y}}{F_{0}} \left[ 1- 1\slash I_{0}\lb R\slash\lambda_{D}\rb \right]^{-1} \rb \nonumber\\
	&\sim\begin{cases}
		\tau f^{-1}\lb \dfrac{\sigma_{Y}}{F_{0}}\rb =\text{const.}& \qquad \text{if} \;\; R\gg\lambda_{D} \\
		&\\
		\tau f^{-1}\lb\dfrac{4\sigma_{Y}}{F_{0}}\cdot\dfrac{\lambda_{D}^{2}}{R^{2}} \rb & \qquad \text{if} \;\; R\ll\lambda_{D}, \end{cases}
\end{align*}
where $I_{0}(z)$ is the modified Bessel function of the first kind and $ I_{0}(z)  \sim e^{2z}\slash \sqrt{ z} $ as $z\to \infty$, and $I_{0}(z) \simeq 1+z^{2}\slash 4$ near $z=0$. 
Therefore, this form is consistent with the result of the dimensional analysis.

\end{document}